\documentclass{PoS}

\title{Preliminary results of a WIMP search with EDELWEISS-II cryogenic detectors}

\ShortTitle{Preliminary results of a WIMP search with EDELWEISS-II cryogenic detectors}

\author{\speaker{Eric Armengaud}\\
        IRFU/SPP, CEA-Saclay, 91191 Gif-sur-Yvette, France\\
        E-mail: \email{eric.armengaud@cea.fr}}

\abstract{The EDELWEISS-II experiment uses cryogenic heat-and-ionization detectors in order to detect the rare interactions from possible WIMP dark matter particles on Germanium nuclei. Recently, new-generation detectors with an interleaved electrode geometry were developped and validated, enabling an outstanding rejection of gamma-rays and surface interactions. We present here preliminary results of a one-year WIMP search carried out with ten of such detectors in the Laboratoire Souterrain de Modane. A sensitivity to the spin-independent WIMP-nucleon cross-section of $5\times 10^{-8}$ pb was achieved using a 322 kg$\cdot$days effective exposure. We also present the current status of the experiment and prospects to improve the present sensitivity by an order of magnitude in the near future.}

\FullConference{Identification of Dark Matter 2010-IDM2010\\
		July 26-30, 2010\\
		Montpellier France}

\begin{document}

\section{The EDELWEISS-II set-up and detectors}
The existence of Weakly Interacting Massive Particles (WIMPs) is a likely explanation for the various observations of a dark matter component from the largest scales of the Universe to galactic scales~\cite{reviews}. EDELWEISS-II is a low-background, cryogenic experiment aiming at the direct detection of local WIMPs which may constitute the dark matter of our Milky Way. The elastic collisions of WIMPs on ordinary matter target nuclei generate low-energy deposits ($\leqslant 100$ keV) with an exponential-like spectrum, and a very low interaction rate which is currently constrained at the level of $\lesssim 1$ event/kg/year. Detecting these events requires an ultralow radioactivity environment as well as detectors with a low energy threshold and an active rejection of the residual backgrounds~\cite{rev-exp}.

The main EDELWEISS-II setup is located at the Laboratoire Souterrain de Modane (LSM), where the 4800~m water-equivalent rock overburden reduces the cosmic muon flux down to $\sim 5 \,\mu$/day/m$^2$. Germanium detectors are hosted within a reversed geometry dilution cryostat which may host up to 40~kg of detectors down to $\sim$20~mK. A 20~cm thick lead shield surrounds the detectors, with its inner parts made of roman lead, in order to attenuate the external $\gamma$ radioactivity. A 50~cm thick polyethylen shielding protects detectors against the external flux of fast neutrons. A muon veto made of plastic scintillators with a coverage of more than 98\% tags neutrons produced by the residual flux of muons which have been interacting mostly in the lead shield. Additional background monitorings are achieved by a Radon level detector near the cryostat, measurements of the thermal neutron flux inside the shielding with a $^3$He-gas detector, and studies of muon-induced neutrons with a Gd-loaded liquid scintillator outside the shielding.

\begin{figure}[!ht]
  \center{\includegraphics[width=.8\textwidth]{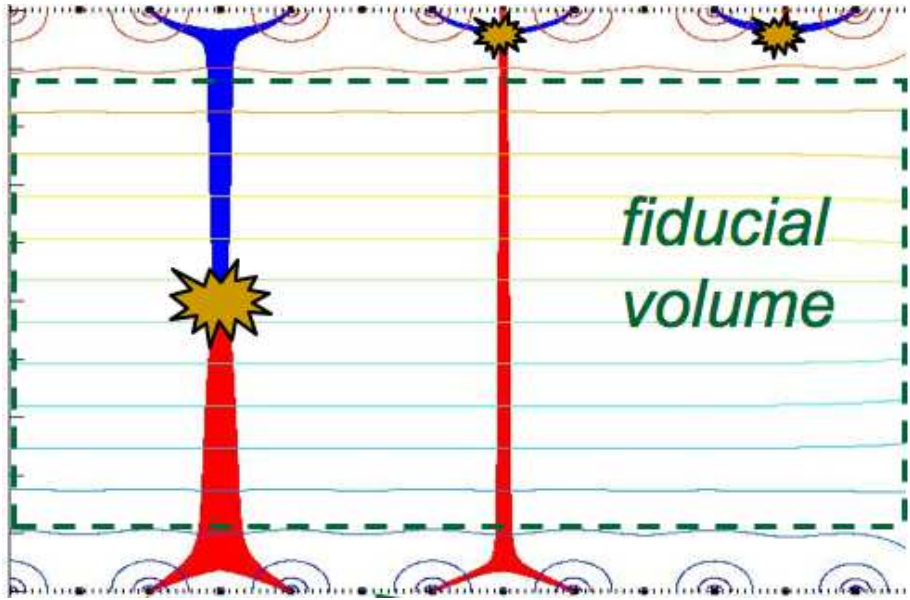}}
  \caption{Sectional view illustrating the principle of an ID detector. The interleaved electrodes on the top and bottom of the Germanium crystal create an electric field whose equipotential lines are represented in thin lines. The charge propagation is represented under very simplified assumptions in blue (electrons) and red (holes) for three different interaction positions within the crystal. The distribution of the collected charges in the different electrodes enables to discriminate interactions taking place within the fiducial volume.\label{fig:id}}
\end{figure}

EDELWEISS detectors are ultrapure germanium crystals equipped with a dual heat and ionization measurement in order to discriminate $\gamma$-induced electronic recoils from potential WIMP-induced nuclear recoils, a technology with proven rejection efficiency since the beginning of the 2000s~\cite{edw1}. The heat sensors are NTD thermometers glued on the surface of each detector, while the ionization is measured using electrodes polarized at low voltages (a few volts). The ionization yield for nuclear recoils is well-measured and is a factor $\sim 3$ lower with respect to electron recoils in the energy region of interest, enabling a complete event-by-event discrimination for the bulk of $\gamma$-ray radioactivity. Still, the charge collection is incomplete and difficult to control when an interaction takes place near the surface of the detector. In particular, the local $\beta$ radioactivity from residual $^{210}$Pb, a daughter of Radon which is present on all surfaces, generates such surface events. Those cannot be discriminated from potential WIMP-induced signals with their ionization yield. In order to reject these events, new-generation detectors called "ID" (InterDigit) were developped. The principle of these detectors is shown on Fig.~\ref{fig:id} : a set of interleaved electrodes forming concentric rings modifies the electric field topology near the crystal surface, and therefore the repartion of charges induced by near-surface events. This allows to tag specifically near-surface interactions and to define a well-controlled fiducial volume for each detector. Note that this strategy to reject surface backgrounds is different from the one used within the CDMS experiment~\cite{cdms}, which relies on the use of more complex phonon sensors.

\begin{figure}[!ht]
  \center{\includegraphics[width=.9\textwidth]{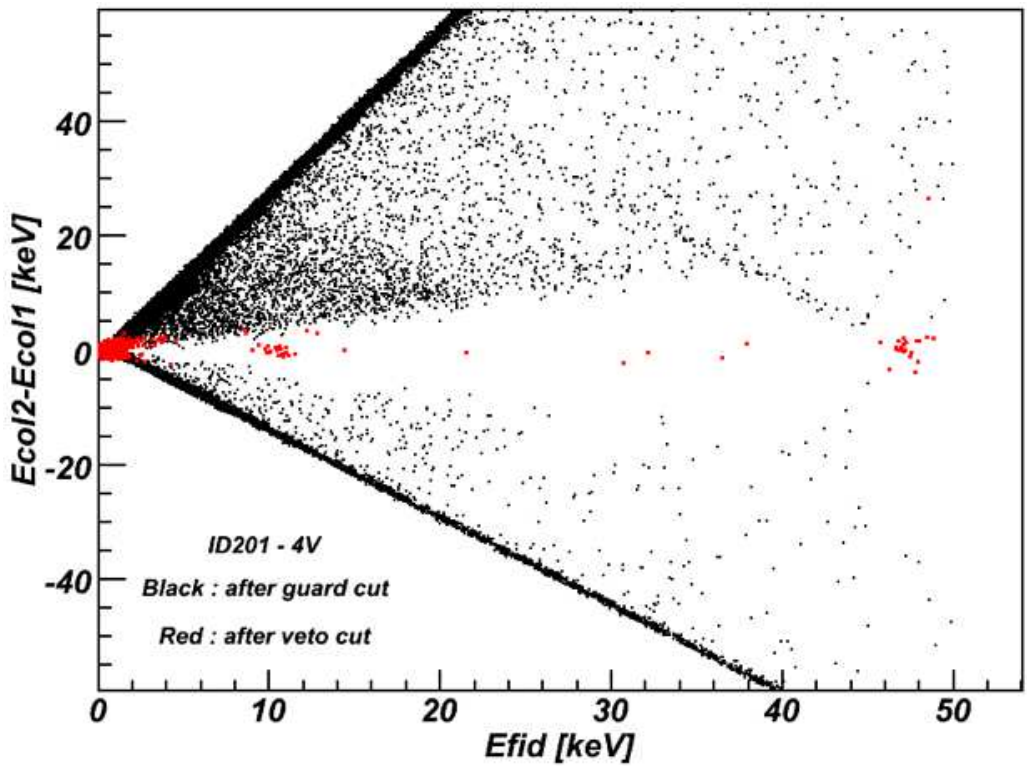}}
  \caption{Difference of signals between the two collecting electrodes as a function of the energy for a calibration with two $^{210}$Pb sources facing each side of a 200~g ID detector with two different rates.  The majority of interactions with $E_{1} \neq E_{2}$ are $\beta$ events. The $\gamma$-ray line at 46 keV is observed, with interactions occuring at the boundary of the fiducial volume. The 10~keV cosmogenic $\gamma$-ray doublet in the fiducial volume is also observed.\label{fig:charge}}
\end{figure}

The discrimination against surface events for ID detectors was experimentally proven at a level of $\sim 10^{-5}$ in~\cite{id-paper} using $\beta$-ray sources placed in front of a 200~g ID bolometer. Fiducial interactions are selected down to low energies by requiring both a perfect charge balance between the signals of the top and bottom collecting electrodes, as shown in Fig.~\ref{fig:charge}, and the absence of charge deposits on the field shapping electrodes. This dual rejection provides a strong redundance, and enables the detector operation even when charges cannot be read on one of the field shapping electrodes. The rejection works quite independently of the intensity of the applied voltages.

The fiducial volume of ID detectors may be estimated using a simple electrostatic modelling of the detectors, and has been also measured in real WIMP-search conditions using the 9.0~keV and 10.3~keV $\gamma$-ray lines from the $^{68}$Ge and $^{65}$Zn isotopes, homogeneously distributed within the crystals due to cosmogenic activation. Comparing the intensity of these lines before and after the fiducial selection provides a measurement of the fiducial mass : for 400~g ID detectors, this mass is $166\pm6$g. Simulations show that it is primarily fixed by the guard regions that are equipped with independent plain electrodes on the edges of the crystals.

\section{A search for WIMPs with ID detectors}

A WIMP search was carried out using ten 400~g ID detectors within the EDELWEISS-II setup, mostly  between April 2009 and May 2010. Here we present a preliminary analysis of the full data set, using nine out of the ten bolometers and corresponding roughly to a doubling of the exposure with respect to the first 6 months of WIMP search which were already published in~\cite{wimp-6months} .

Stable cryogenic conditions at 18~mK were maintained over the whole year without any major interruption. Around 90\% of the electronics channels were working, enabling the use of nine detectors over the ten installed in the cryostat for a first blind WIMP search. External gamma-ray and neutron calibrations were carried out for all detectors. The data was processed using two independent reconstruction pipelines, both of them making use of optimal filtering in order to adapt to the different noise conditions encountered. The average baseline resolutions were of $\sim$ 1.2 keV FWHM for heat channels and $\sim 0.9$ keV FWHM for ionizations. Baseline noise measurements are used in order to automatically select good periods of data taking, with an 80\% efficiency. A $\chi^2$ cut is also applied to reject misreconstructed events. WIMPs are then searched among fiducial events, using the ionization yield to discriminate gamma-rays with a 99.99\% rejection efficiency. In order to reject neutron-induced recoils, events in coincidence between bolometers and also with the muon veto are also rejected. Furthermore, a WIMP search energy threshold was also set a priori at 20 keV in this analysis, so that the search efficiency is independent of the energy. After all cuts, an effective exposure of 322 kg$\cdot$days is obtained. This analysis procedure is strictly identical to the one published in~\cite{wimp-6months}.

\begin{figure}[!ht]
  \includegraphics[width=.49\textwidth]{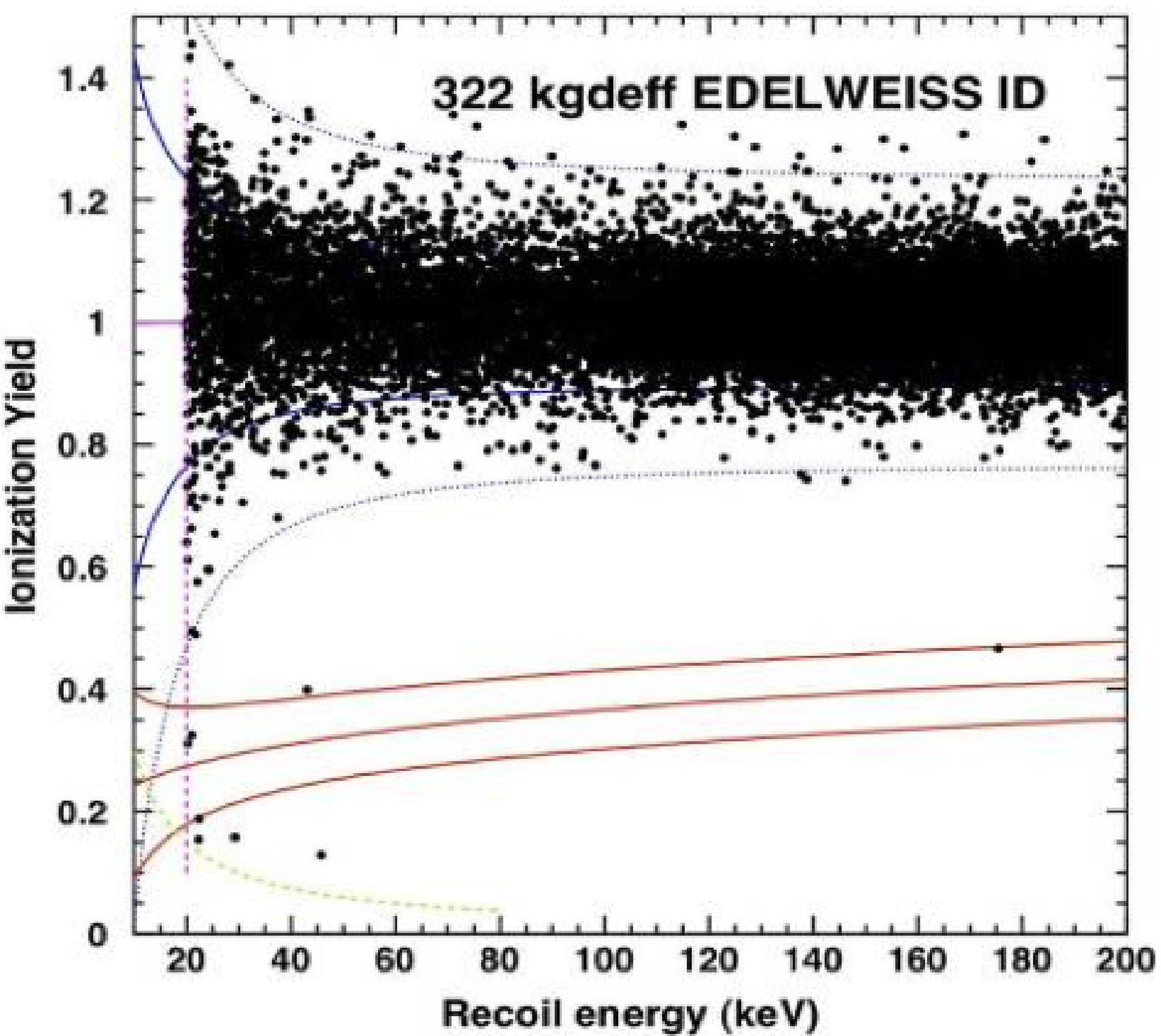}
  \includegraphics[width=.49\textwidth]{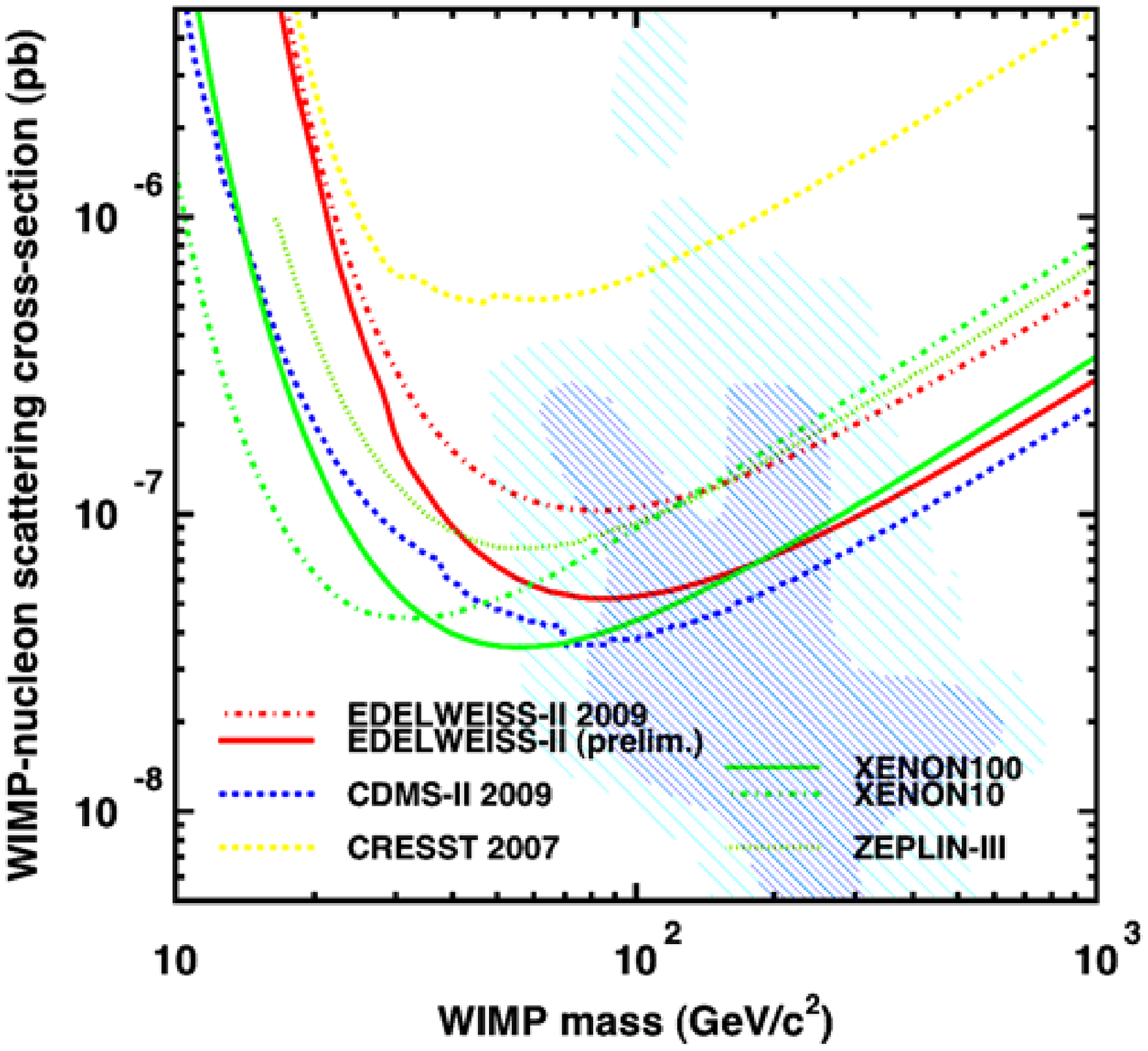}
  \caption{Left : Ionization yield vs energy for all fiducial interactions taking place in a WIMP search corresponding to a 322 kg$\cdot$days exposure. Right : The corresponding limit on the WIMP-nucleon cross-section (continuous red line) is compared to other experiments.\label{fig:results}}
\end{figure}

The WIMP-gamma discrimination diagram obtained after the fiducial cut is presented on Fig.~\ref{fig:results} (left). Four events are observed in the band, but with a large gap with no event between 23 and 175 keV. Using the standard Yellin prescription~\cite{yellin}, this enables to improve the EDELWEISS sensitivity by a factor of 2 with respect to the published limit, as shown on Fig.~\ref{fig:results} (right). A limit on the spin-independent WIMP-nucleon cross-section of $5\times 10^{-8}$ pb is set for a WIMP mass of 80 GeV. A final analysis of this data set is under way, which aims mostly at optimizing the cumulated exposure using our improved knowledge of these new detectors.

From the Fig.~\ref{fig:results} (left), first hints of a residual background start to appear in the data from these detectors : three events are present in the nuclear recoil band near the analysis threshold, between 20 and 23 keV, while two other outliers are present near the nuclear recoil band at higher energies. Note that all these events are well-reconstructed, and comfortably above the noise level of the detectors. Background studies are currently ongoing to fully understand their origin. Preliminary upper limits may be derived for the known residual gamma, beta and neutron backgrounds, using calibrations, radioactivity measurements as well as simulations. Overall, less than 1.6 events (90\% CL) of known origin are expected for this WIMP search. 

\section{Current status and prospects}

\begin{figure}[!ht]
  \center{\includegraphics[width=.7\textwidth]{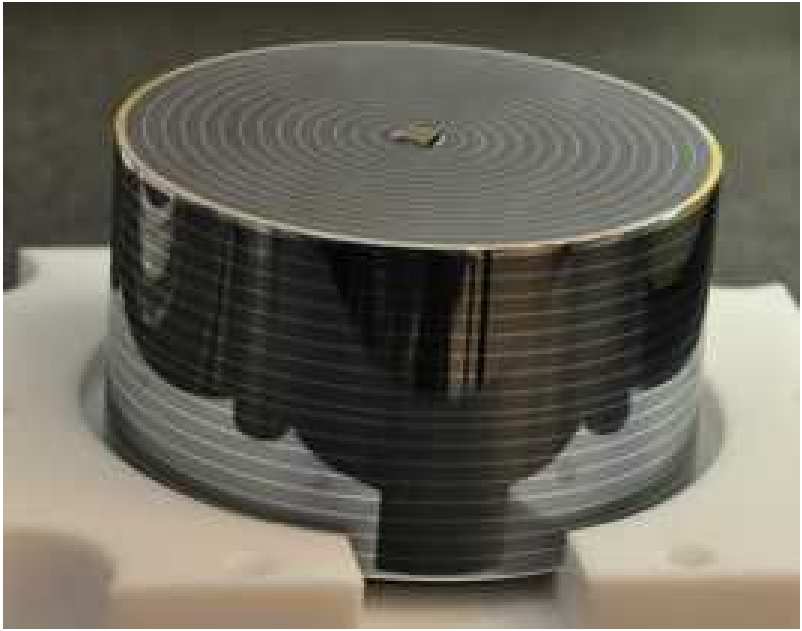}}
  \caption{Picture of an 800~g Full InterDigit bolometer.\label{fig:fid800}}
\end{figure}

Optimized InterDigit detectors are being installed and commissionned at the LSM. The interleaved set of electrodes now surrounds also the edges of the crystal, hence their name Full InterDigit (FID). A first series of four 800~g and two 400~g FIDs were built, one of them been shown on Fig.~\ref{fig:fid800}. By using both the FID design and an unprecedented mass of 800~g, this will increase significantly the fiducial mass for a given detector. The 800~g detectors currently have six charge readout channels and two NTD sensors, enabling an additional redundancy. Several surface treatments are also under study to increase the surface event rejection. Extensive $\gamma$-ray calibrations are already being carried out at the LSM with these detectors.

After the validation of these "FID800" detectors, it is planned to upgrade several parts of the EDELWEISS-II setup (shieldings, cryostat, cabling) to reduce the gamma and neutron backgrounds and lower the energy thresholds. Then 40 such detectors will be installed allowing to reach a 3000~kg$\cdot$days exposure within a few months in 2012, with a potential WIMP-nucleon cross-section sensitivity at the level of $5\times 10^{-9}$ pb.

As a conclusion, the EDELWEISS-II collaboration has carried out a one-year WIMP search with new-generation ID detectors. The achieved sensitivity, of $5\times 10^{-8}$~pb for a WIMP mass of 80~GeV  is at the same level than the recently published CDMS and XENON100 sensitivities to the standard neutralino-like WIMP models. Optimized detectors are now on the way to be validated in order to improve the current sensitivity by an order of magnitude within the coming years.

\end{document}